\pdfoutput=1
\documentclass[10pt,a4paper,twocolumn]{article}
\usepackage[top=1in, bottom=1in, left=1in, right=1in]{geometry}
\usepackage[utf8]{inputenc}
\usepackage{bm}
\usepackage{float}
\usepackage{amsmath}
\usepackage{amsfonts}
\usepackage{amssymb}
\usepackage{makeidx}
\usepackage{graphicx}
\usepackage{lmodern}

\title{{A prescription for transforming polarization states of light using two quarter waveplates}}

\author{B. Radhakrishna\footnote{brkrishna@igcar.gov.in}\\ 
Materials Science Group,\\
Indira Gandhi Centre for Atomic Research, \\
HBNI, Kalpakkam, 603102, India.\\
\normalsize{(Email: brkrishna@igcar.gov.in)}}
\date{}
\begin{document}
\twocolumn[
 \begin{@twocolumnfalse}
\maketitle
\begin{abstract}
It is well known that state transformation from one polarization state  to another  can be achieved with a minimal gadget consisting of two quarter waveplates. A constructive, geometric approach is presented, which provides a direct prescription for the fast axis setting of the waveplates to transform any completely polarized states of light to any other, including orthogonal ones.   
\\
\\
\end{abstract}
 \end{@twocolumnfalse}
]
\section{Introduction}
Manipulating polarization states of light with a minimum number of optical elements is an important requirement in the field of polarization optics.  Implementation of quantum walks using linear optics\cite{Qwalk}, measurement of quantum correlations of optical systems\cite{GS_Agarwal}, quantum state tomography, determination of  Mueller matrix of arbitrary optical materials\cite{Mueller}, measurement of the Pancharatnam phase \cite{Pancharatnam-RPS, pancharatnam_Dezela} and the  generation of mixed states\cite{MixedQ, partially} are some typical examples where such gadgets are being used in the present scenario. In all these applications, the general requirement is to realize a unitary state transformations with the help of a minimal number of linear optical elements.
\\
\\
Theoretically, any two polarization states of light having the same degree of polarization can be transformed into one another using an appropriate SU(2) operation. It has been shown in the literature, that the mimum requirement for realizing the complete set of  SU(2) transformations are a combination of two quarter waveplates(QWPs) and one half waveplate(HWP)\cite{RSNM-1, RSNM-2}. However, for transforming any two polarization states of light having the same degree of polarization, the complete set of  SU(2) transformations may not be required. Infact, a subset of SU(2) transformations generated from a combination of two QWPs will be sufficient\cite{Damask, SGReddy}. In this article, we provide a mathematical treatment to validate it, by considering the transformation between any two completely polarized state of light (CPSL). This result is generalizable to other degrees of polarization as well. Here, we also provide an algorithm to determine the required orientation of two QWPs for mutually transforming any two CPSL. With our approach, we overcome the limitations of an earlier approach\cite{De-Zela}, which concluded that two QWPs are insufficient to deal with mutually orthogonal states.
\\ 
\noindent This article is organized as follows: in section 2, we describe our approach for validating the above result. In section 3, we discuss the results obtained through our approach and finally, we conclude this article in section 4. 
\section{Assignment of symmetry operations}
\begin{figure}
\centering \includegraphics[width=0.5767\linewidth]{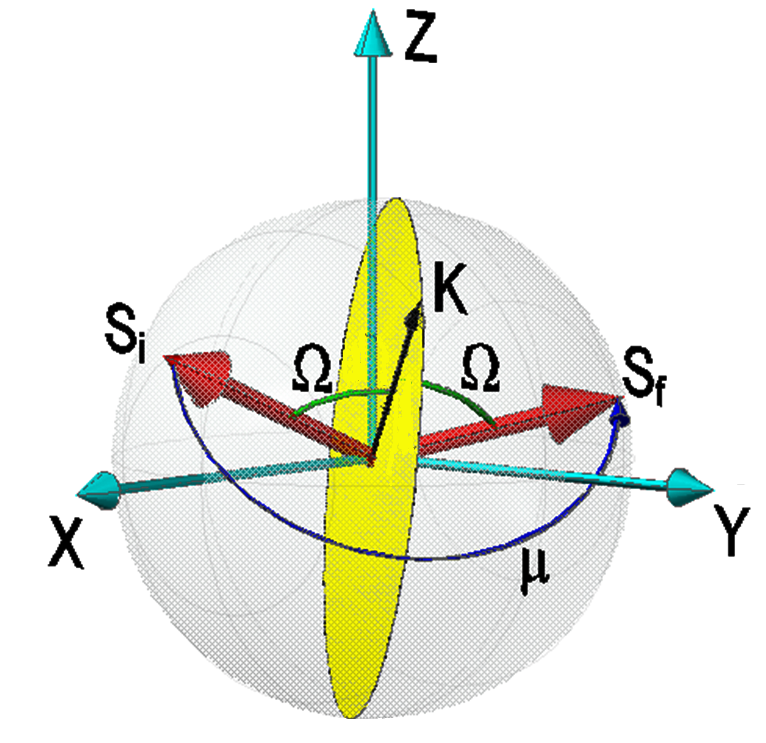}
\caption{\small{[Color online] CPSL $\bm{S_{i}}$ is rotated about the axis $\bm{k}$, by an angle $\mu$ to reach another CPSL $\bm{S_{f}}$. Rotation axis $\bm{k}$ point radially outward in the mid plane(yellow region) of $\bm{S_{i}}$ and $\bm{S_{f}}$.}}
\label{Axis} 
\end{figure}

\begin{figure}
\centering \includegraphics[width=0.567\linewidth]{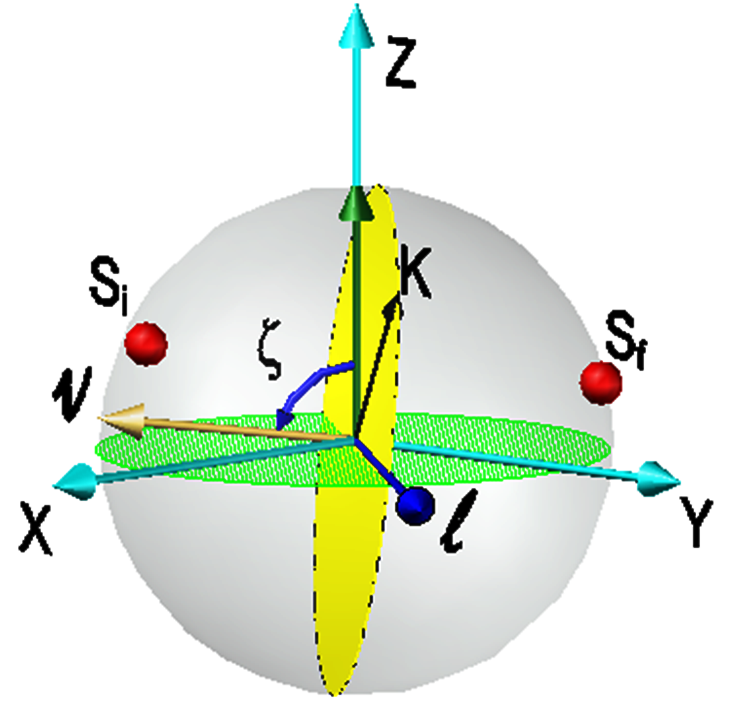}
\caption{\small{[Color online] Mid-plane of $\bm{{S}_{i}}$, $\bm{{S}_{f}}$ is $\bm{v}=\bm{{S}_{i}}-\bm{{S}_{f}}$ (yellow region). On rotating the unit circle in in the plane normal to $\bm{z}$ (green region) about $\bm{l}=\bm{z}\times\bm{v}$ by an angle $\zeta=\cos^{-1}(\bm{z\cdot v})$, the unit circle in the mid-plane $\bm{v}$ (yellow region) is obtained.}}
\label{z-to-v} 
\end{figure}

The issue of transforming any CPSL to any other CPSL can be addressed on the Poincare sphere. Poincare sphere is a widely used construction for describing the polarization states of light, where various polarization states are mapped onto a solid sphere of unit radius\cite{DH-Goldstein, Collet}. A CPSL is mapped on to the surface of the Poincare sphere and it has three real coordinates $\bm{S}=(s_{x}, s_{y}, s_{z})$ satisfying the condition ${s_{x}}^{2}+{s_{y}}^{2}+{s_{z}}^{2}=1$. $\bm{S}$ is called the Bloch vector of a polarization state, which can be described in terms of spherical-polar coordinates $\theta$ and $\phi$ as:
\begin{equation}\label{theta-phi}
\bm{S}(\theta,\phi)=\begin{pmatrix}
\cos \phi\, \sin \theta \\
\sin \phi\, \sin \theta \\
\cos \theta 
\end{pmatrix}
\end{equation}
On the Poincare sphere, the task of transforming any initial CPSL to any final CPSL, is equivalent to transforming their respective Bloch vectors $\bm{S_{i}}(\theta_{i},\phi_{i})$ to $\bm{S_{f}}(\theta_{f},\phi_{f})$. Such norm preserving transformations are realized as by rotating $\bm{S_{i}}$ about a suitable axes $\bm{k}$ by an appropriate angle $\mu$ to reach $\bm{S_{f}}$ as shown in figure \ref{Axis}.
\noindent For these transformations, all possible rotation axes point radially outward in the mid-plane of $\bm{S_{i}}$ and $\bm{S_{f}}$ whose plane normal is given by $\bm{v}(\theta,\phi)=\bm{{S}_{i}}-\bm{{S}_{f}}$, where $\bm{v}(\theta,\phi)$ is an another Bloch vector. The functional form of the various rotation axes in the mid-plane $\bm{v}(\theta,\phi)$ are obtained by parameterizing the unit circle in the plane $\bm{v}(\theta,\phi)$ denoted by $\bm{k}_{\bm{v}(\theta,\phi)}(\nu)$, where $\nu$ is a parameter. An explicit form of $\bm{k}_{\bm{v}(\theta,\phi)}(\nu)$ is obtained by using a parametric form of an unit circle in any known plane and the Rodrigues operator\cite{CM-Goldstein}. A parametric form of the unit circle centered at the origin in the plane normal to $\bm{z}$ is given by:
\begin{equation}
\nonumber
\bm{k}_{\bm{z}}(\nu)=\begin{pmatrix}
\cos\nu \\ \sin\nu \\ 0\end{pmatrix},\quad \text{where}\, \nu\in(0,2\pi). 
\end{equation} 
The Rodrigues operator $\mathcal{\hat{R}}(\bm{l},\zeta)$ dictates the result of rotating any vector (say $\bm{z}$) about any unit axis (say $\bm{l}$) by an angle $\zeta$ given by:
\begin{equation}\label{Rodrigues-org}
\small
\begin{split}
\bm{v}\scriptstyle(\theta,\phi)& =  \mathcal{\hat{R}}(\bm{l},\zeta) \, \bm{z}\\
\bm{v} & =  \cos\zeta\ \bm{z}+(1-\cos\zeta)(\bm{l}\bm{\cdot}\bm{z}){\bm{l}}+\sin\zeta({\bm{l}}\times\bm{z})\\
\end{split}
\end{equation}
The intended parametric form of a unit circle in the plane normal to $\bm{v}\scriptstyle(\theta,\phi)$ and hence the parametric form of rotation axes $\bm{k}_{\bm{v}(\theta,\phi)}(\nu)$ is obtained by (see figure \ref{z-to-v}):
\begin{equation}\label{Rodrigues}
\small
\begin{split}
\bm{k}_{\bm{v}(\theta,\phi)}(\nu) & =  \mathcal{\hat{R}}(\bm{l},\zeta)\bm{k}_{\bm{{z}}}(\nu)\\[1.5ex]
&  =  \begin{pmatrix}
\cos\theta\cos\nu - \left(1-\cos\theta\right)\sin\textstyle{(\nu-\phi)} \,\sin\phi \\[.25ex]
\cos\theta\sin\nu+\left(1-\cos\theta\right)\sin(\nu-\phi) \,\cos\phi \\[.25ex]
-\sin\theta \cos(\nu-\phi) 
\end{pmatrix}\\
\end{split}
\end{equation}
where $\bm{l}=\bm{z}\times\bm{v}$ and $\zeta=\cos^{-1}(\bm{z\cdot v})=\theta$, obtained from eqn. \ref{theta-phi}.\\
\\
\noindent In an earlier approach\cite{De-Zela}, the rotation axis $\bm{k}$ was written down as a linear combination of three vectors namely: $\bm{S_{i}}$, $\bm{S_{f}}$ and $\bm{S_{i}}\times\bm{S_{f}}$. But for orthogonal states  $\bm{S_{f}}=-\bm{S_{i}}$ and the chosen basis set is no longer linearly independent. This led them to conclude that the two QWP gadget is insufficient for transforming orthogonal states. The result stems from the limitations of the algorithm and the basis choice adopted therein, and not due to the deficiency of the gadget itself. The present work parametrizes the rotation axis in a manner that avoids this problem.\\
\\
\noindent In order to validate the transformation from any CPSL $\bm{S_{i}}$ to any other CPSL $\bm{S_{f}}$, we first determine the angle $\mu$ required for rotating $\bm{S_{i}}$ about the rotation axes $\bm{k}_{\bm{v}(\theta,\phi)}(\nu)$ to reach $\bm{S_{f}}$. Then, we calculate the rotation angle $\mu_{q}$ achievable from the two QWP gadget about the same axes. Finally, we look for rotation axes $\bm{k}_{\bm{v}(\theta,\phi)}(\nu)$, about which the required rotation angle is achievable from two QWPs.
\\
\\
The rotation angle $\mu$ required in transforming $\bm{S_{i}}(\theta_{i},\phi_{i})$ to $\bm{S_{f}}(\theta_{f},\phi_{f})$, about any unit axis $\bm{k}_{\bm{v}(\theta,\phi)}(\nu)$ is determined from Rodrigues operator as:
\begin{equation}\label{Si-R-Sf}
\small
\bm{S_{f}} = \cos\mu\ \bm{S_{i}}+(1-\cos\mu)\,({\bm{k_{v}}}\bm{\cdot}\bm{S_{i}})\,{\bm{k_{v}}}+\sin\mu\, ({\bm{k_{v}}}\times\bm{S_{i}})
\end{equation}
As the rotation axes $\bm{k_{v}}(\nu)$ lies in the mid-plane of $\bm{S_{i}}$ and $\bm{S_{f}}$ (see figure \ref{Axis}), hence: 
\begin{equation}\label{kdot-S}
\bm{k_{v}}(\nu)\bm{\cdot}\bm{S_{i}} =  \bm{k_{v}}(\nu)\bm{\cdot}\bm{S_{f}} =\cos\Omega
\end{equation}
From \ref{Si-R-Sf} and \ref{kdot-S}, angle $\mu$ required about the rotation axis $\bm{k_{v}}(\nu)$ in transforming $\bm{S_{i}}$ to $\bm{S_{f}}$ is 
\begin{equation}\label{mu-reqd}
\cos\mu\,\big( 1-\cos^2\Omega \big)-\bm{S_{i}}\cdot\bm{S_{f}} + \cos^2\Omega=0
\end{equation}

\noindent For identifying the rotation angle $\mu_{q}$ achievable about any axis using two QWP gadget, we 
study the Cayley-Klein parameters\cite{CM-Goldstein} of the gadget. Cayley-Klein parameters $(e_{0},\,\bm{e})$ of any SU(2) transformation, are the coefficients of that transformation expressed as a linear combination of Pauli-spin matrices $\bm{\sigma}$ and the identity matrix as basis.
\begin{equation}\label{qwp-mu}
\begin{split}
\bm{U}=&\exp\left(-i{\textstyle \frac{\mu_{q}}{2}}\bm{k\cdot\sigma}\right) \in\text{\;SU(2)}\\
=&\cos\left({\textstyle \frac{\mu_{q}}{2}}\right) \bm{I}-i\sin\left({\textstyle {\frac{\mu_{q}}{2}}}\right)(\bm{k}\cdot\bm{\sigma})
\end{split} 
\end{equation}
Cayley-Klein parameters of $\bm{U}$ are:
\begin{equation}\label{2qwp-CK}
\begin{split}
e_{0}=\cos\left({\textstyle \frac{\mu_{q}}{2}}\right); \quad e_{x}=\sin\left({\textstyle \frac{\mu_{q}}{2}}\right)k_{x}\\
e_{y}=\sin\left({\textstyle \frac{\mu_{q}}{2}}\right)k_{y};\quad e_{z}=\sin\left({\textstyle \frac{\mu_{q}}{2}}\right)k_{z}
\end{split}
\end{equation}

\begin{figure}
\centering \includegraphics[width=0.87\linewidth]{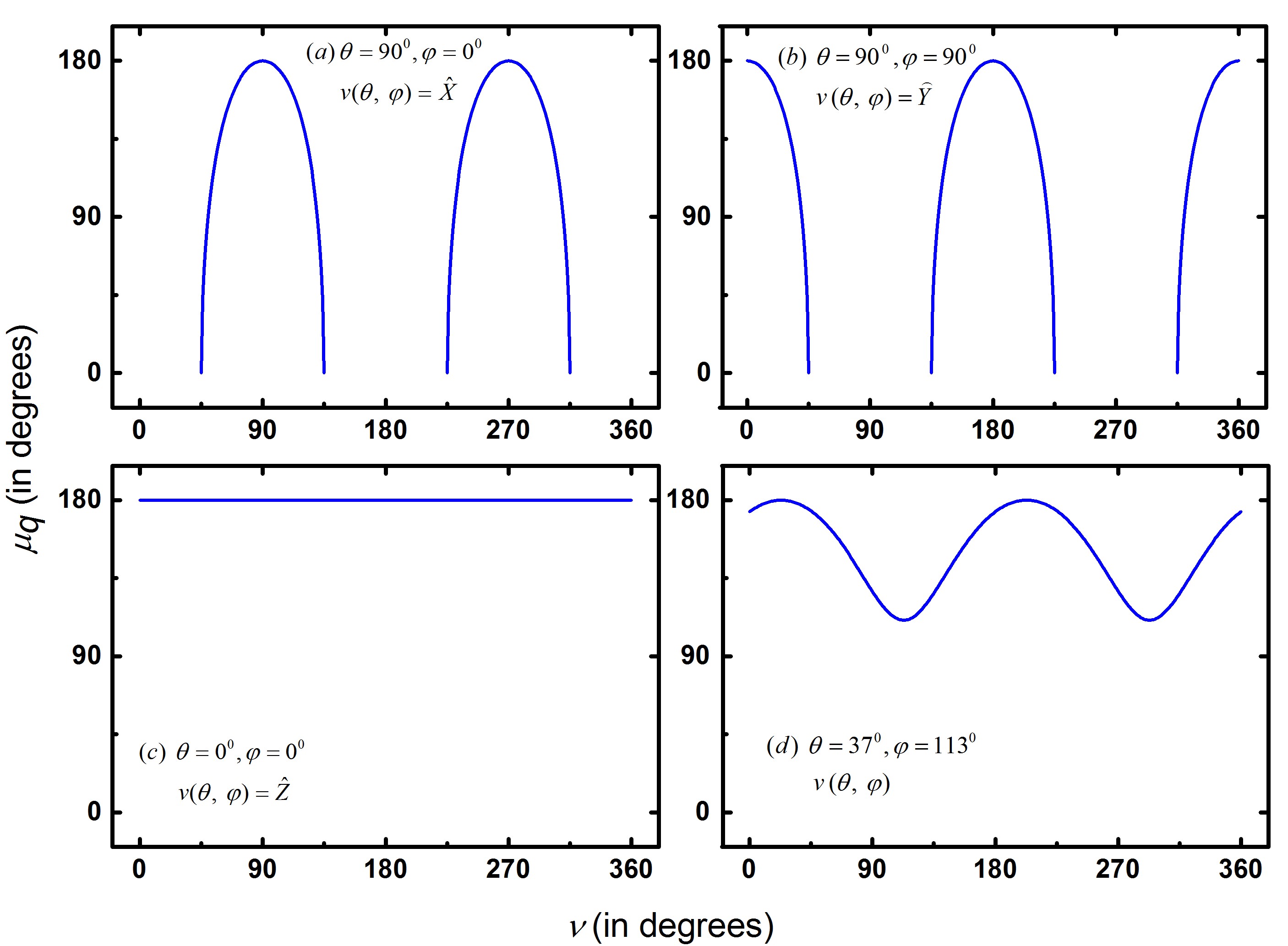}
\caption{\small{[Color online] In different mid-planes $\bm{v}(\theta,\,\phi)$, the rotation angle $\mu_{q}$ achievable from a combination of two QWPs about parameter $\nu$ of the rotation axes $\bm{k}_{\bm{v}(\theta,\,\phi)}(\nu)$ is shown. $\textstyle{(a)\,\theta=90^{0},\,\phi=0,\,\bm{v}(\theta,\phi)=\bm{X}}$ $\textstyle{(b)\,\theta=90^{0},\,\phi=90^{0},\,\bm{v}(\theta,\phi)=\bm{Y}}$
$\textstyle{(c)\,\theta=0,\,\phi=0,\,\bm{v}(\theta,\phi)=\bm{Z}}$ $\textstyle{(d)\,\theta=37^{0},\,\phi=113^{0},\,\bm{v}(\theta,\phi)}$.}}
\label{all4-in-one} 
\end{figure}

\noindent The operator $\bm{U}$ describes the rotation about the unit axis $\bm{k}$ by an angle $\mu_{q}$. Here, we are using the convention $\bm{\sigma_{x}}=\left(\begin{smallmatrix}1 & 0\\
0 & -1
\end{smallmatrix}\right)$, $\bm{\sigma_{y}}=\left(\begin{smallmatrix}0 & 1\\
1 & 0
\end{smallmatrix}\right)$, $\bm{\sigma_{z}}=\left(\begin{smallmatrix}0 & -i\\
i & 0
\end{smallmatrix}\right)$. Cayley-Klein parameters of every SU(2) matrix satisfies
\begin{equation}\label{quatrenions}
e_{0}^2+e_{x}^2+e_{y}^2+e_{z}^2 = 1 
\end{equation}
From eqn. \ref{2qwp-CK}, it is evident that the above condition is independent of the rotation axis $\bm{k}$ and the angle $\mu_{q}$. In other words, rotation about any unit axis $\bm{k}$ by any angles $\mu_{q}$ are achievable from entire SU(2) transformations.\\
\\
The Cayley-Klein parameters of a two QWP gadget $\bm{U_{QQ}} =  \bm{Q}(\eta_{2})\bm{Q}(\eta_{1})$ is 
\begin{equation}\label{qq-alpha-beta}
\begin{split}
e_{0}=\sin^{2}\beta; \quad e_{x}=\cos\alpha\cos\beta\\
e_{y}=\sin\alpha\cos\beta;\quad e_{z}=\sin\beta\cos\beta
\end{split}
\end{equation}
where $\bm{Q}(\eta)$ is the description of a QWP in the basis of Pauli spin matrices, given by:
\begin{equation}\label{qwp_eqn}
\small
\begin{split}
\bm{Q}(\eta)= & \dfrac{1}{\sqrt{2}}\left(\bm{I}-i\cos2\eta\,\bm{\sigma_{x}}-i\sin2\eta\,\bm{\sigma_{y}}\right)\, \in \,\text{SU(2)} \\
\end{split}
\end{equation}
and $\eta$ is the angle between the fast axis of a QWP and the reference axis. We have choosen $\eta_{1}=\frac{\alpha+\beta}{2},\,\eta_{2}=\frac{\alpha-\beta}{2}$. It may be noted that two QWP gadget along with the condition \ref{quatrenions} also satisfies:
\begin{equation}\label{quat-2qwp}
e_{0}+e_{x}^2+e_{y}^2=1
\end{equation}
Combining eqns. \ref{quatrenions}, \ref{quat-2qwp}:
\begin{equation}
e_{0}^2-e_{0}+e_{z}^2=0\\
\end{equation} 
From eqns. \ref{Rodrigues} and \ref{2qwp-CK},
\begin{equation}\label{spl-achievble}
\cos^2\left({\textstyle \frac{\mu_{q}}{2}}\right)-\cos\left({\textstyle \frac{\mu_{q}}{2}}\right)+\sin^2\left({\textstyle \frac{\mu_{q}}{2}}\right){\sin^2\theta \cos^2(\nu-\phi)}=0
\end{equation} 
The above equation describes the rotation angle $\mu_{q}$ achievable about the rotation axis $\bm{k}_{\bm{v}(\theta,\phi)}(\nu)$ using two QWPs gadget. This equation is quadratic in $\cos\left({\textstyle \frac{\mu_{q}}{2}}\right)$ and its non-trivial solution is 
\begin{equation}\label{achievable}
\cos\mu_{q}  =\, 2\bigg(\dfrac{A}{1-A}\bigg)^2-1 
\end{equation}
where $A={\sin^2\theta \cos^2(\nu-\phi)}$.
\\
\\
In order to get some insight as to the restrictions on the attainable transformations with a combination of two QWPs, we give some illustrative examples. We study the variation of the rotation angle $\mu_{q}$ achievable from two QWPs as a function of the rotation axes parameter $\nu$ in a certain mid-planes $\bm{v}(\theta,\phi)$ and it is shown in figure \ref{all4-in-one}. We notice that, in some mid-planes: (i) about certain axes, rotations are forbidden (see figure \ref{all4-in-one}$a$, \ref{all4-in-one}$b$), (ii) achievable rotation angles $\mu_{q}$ remain the same, irrespective of the rotation axis (see figure \ref{all4-in-one}$c$), (iii) $\mu_{q}$ lies within a band of angles and varies sinusoidally as a function of the rotation axis (see figure \ref{all4-in-one}$d$). \\

\noindent This variation is seen because, QWPs basically introduces a relative retardance of $\frac{\pi}{2}$ and the fast axis of QWPs are oriented only in the plane perpendicular to the propagation direction of light beam.
\begin{equation}
\begin{split}
\bm{U_{QQ}}=& \exp \left(-i{\textstyle {\frac{\pi}{4}}} \bm{k_{2} \cdot \sigma}\right) \exp \left(-i{\textstyle {\frac{\pi}{4}}} \bm{k_{1} \cdot \sigma}\right) \\
=&\cos\left({\textstyle \frac{\mu_{q}}{2}}\right) \bm{I}-i\sin\left({\textstyle {\frac{\mu_{q}}{2}}}\right)(\bm{k}\cdot\bm{\sigma})
\end{split}
\end{equation} where $\bm{k_{1,2}}$ are the fast axis orientation of two QWPs. $\cos\left({\textstyle \frac{\mu_{q}}{2}}\right)=\frac{1}{2}\big(1-\bm{k_{2}}\cdot\bm{k_{1}}\big)$ and $\sin\left({\textstyle {\frac{\mu_{q}}{2}}}\right)\bm{k}=\frac{1}{2}\Big(\bm{k_{1}}
+\bm{k_{2}}+(\bm{k_{2}}\times\bm{k_{1}})\Big)$. In the two QWP gadget, constraint on the rotation angle $\mu_{q}$ achievable about the rotation axes arises due to the restricted orientation of fast axes $\bm{k_{1,2}}$ in both the QWPs. 

\section{Results and discussion}
\begin{figure}[H]
\centering \includegraphics[width=0.53467\linewidth]{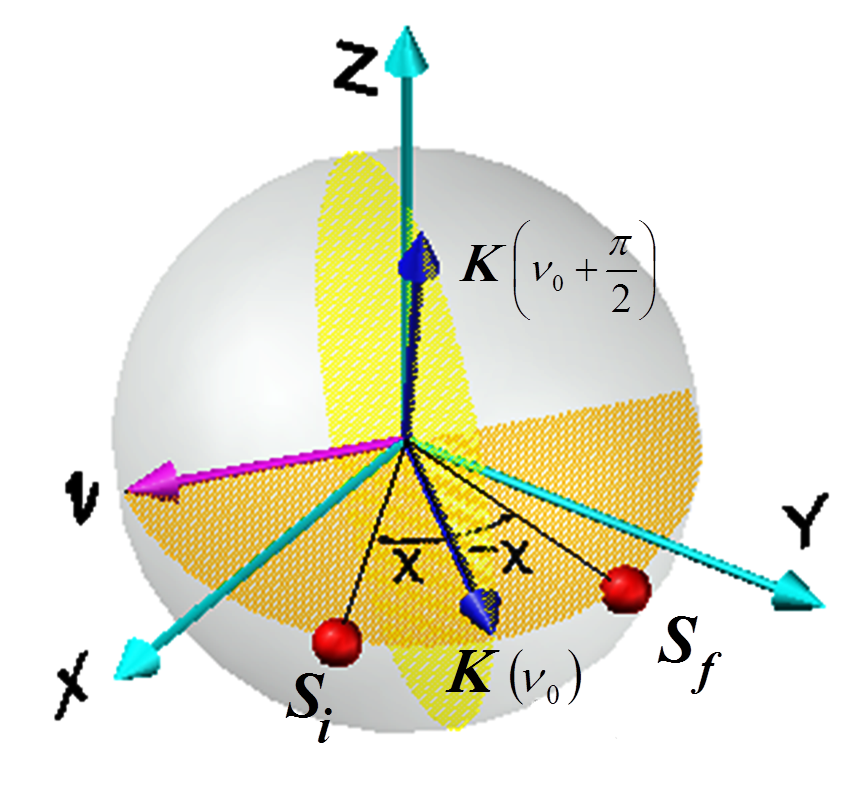}
\caption{\small{[Color online] In the mid-plane $\bm{v}$(yellow region), the rotation axis $\bm{k}_{\bm{v}}(\nu_{0})$ is rotated about its orthogonal axis $\bm{k}_{\bm{v}}(\nu_{0+\textstyle{\frac{\pi}{2}}})$ by angle $\chi$ and $-\chi$. The  resulted vectors on the Poincare sphere are $\bm{S_{i}}$ and $\bm{S_{f}}$ respectively.}}
\label{chi-figure} 
\end{figure}

We shall now discuss the problem of transforming any CPSL to any other CPSL by considering cases of orthogonal states and non-orthogonal states separately.\\
\\
\textbf{(i) Transforming orthogonal states using two QWPs.}\\
On the Poincare sphere, orthogonal states lie at antipodal points and hence the angle $\mu$ required in transforming any pair of orthogonal states about any axis in their mid-plane is always $\pi$. The possibility of achieving rotation angle $\mu_{q}=\pi$ with the two QWPs gadget is analyzed by studying eqn. \ref{spl-achievble}, which simplifies to 
\begin{equation}\label{simplified}
\sin^2\theta \cos^2(\nu-\phi)=0
\end{equation}
For $\theta=0$, it is clear that eqn. \ref{simplified} is satisfied by every $\nu$ and hence, about every axis in the mid-plane $\bm{v}{(0,\phi)}$, rotation angle $\pi$ is achievable. Basically, $\theta=0$ describes the mid-plane whose plane normal is $\bm{z}$. Antipodal points about this mid-plane are $(0,0,1)^T$ and $(0,0,-1)^T$ and these points on the Poincare sphere represents right circularly polarized state (RCP) and left circularly polarized state (LCP) respectively.  Hence, using two QWPs, transformation from RCP to LCP and vice-versa is possible in \textit{infinite ways}, which is a well-established result\cite{SGReddy}.\\
\\
{For $\theta\neq0$}, it is always possible to find $\nu$ satisfying the eqn. \ref{simplified}, irrespective of the mid-plane $\bm{v}(\theta,\phi)$. In other words, for any orthogonal states there always exists a rotation axis $\bm{k}_{\bm{v}(\theta,\phi)}(\nu)$ about which rotation angle $\pi$ is achievable using two QWPs. Hence any orthogonal states are transformable using two QWPs. This result overcomes the limitations of the earlier work\cite{De-Zela}. \\
\\
\textbf{(ii) Transforming non-orthogonal states using two QWPs.}\\
In order to do this, we first identify all possible pairs of CPSL $\bm{S_{i}}$ and $\bm{S_{f}}$, such that their mid-plane is $\bm{v}(\theta,\phi)$. This is done by rotating every axis say $\bm{k}_{\bm{v}(\theta,\phi)}(\nu_{0})$ in the mid-plane $\bm{v}(\theta,\phi)$ about $\bm{k}_{\bm{v}(\theta,\phi)}(\nu_{0}+\textstyle{\frac{\pi}{2}})$, by an equal angles in opposite directions, say $\chi$ and $-\chi$ as shown in figure \ref{chi-figure}. So obtained $\bm{S_{i}}$ and $\bm{S_{f}}$ are restricted to the plane formed by $\bm{v}(\theta,\phi)$ and $\bm{k}_{\bm{v}(\theta,\phi)}(\nu_{0})$. It may be noted that $\bm{v}(\theta,\phi)$, $\bm{k}_{\bm{v}(\theta,\phi)}(\nu_{0})$ and $\bm{k}_{\bm{v}(\theta,\phi)}(\nu_{0}+\textstyle{\frac{\pi}{2}})$ forms a mutually orthogonal basis and in this basis $\bm{S_{i}}$ and $\bm{S_{f}}$ are:
\begin{equation}\label{Si}
\begin{split}
\bm{S_{i}}(\theta,\phi)=&{\mathcal{\hat{R}}}\Big(\bm{k}_{v(\theta,\phi)}(\nu_{0}+\textstyle{\frac{\pi}{2}}), \, \chi\Big)\;\bm{k}_{v(\theta,\phi)}(\nu_{0})\\
= & \cos\chi \,\bm{k}_{\bm{v}(\theta,\phi)}(\nu_{0})\, - \,\sin\chi \,\bm{v}(\theta,\phi)
\end{split}
\end{equation} 

\begin{equation}\label{Sf}
\begin{split}
\bm{S_{f}}(\theta,\phi)=&{\mathcal{\hat{R}}}\Big(\bm{k}_{v(\theta,\phi)}(\nu_{0}+\textstyle{\frac{\pi}{2}}), \, -\chi\Big)\;\bm{k}_{v(\theta,\phi)}(\nu_{0})\\
= & \cos\chi \,\bm{k}_{\bm{v}(\theta,\phi)}(\nu_{0})\, + \,\sin\chi \,\bm{v}(\theta,\phi) 
\end{split}
\end{equation}

\noindent From eqn. \ref{mu-reqd}, angle $\mu$ required in transforming above $\bm{S_{i}}(\theta,\phi)$ to $\bm{S_{f}}(\theta,\phi)$ about axes $\bm{k}_{\bm{v}(\theta,\,\phi)}(\nu)$ is,
\begin{equation}\label{Gen-mu-reqd}
\cos\mu=\frac{\cos2\chi\,-\,\cos^{2}\chi\cos^{2}(\nu-\nu_{0})}{1-\cos^{2}\chi\cos^{2}(\nu-\nu_{0})}
\end{equation}
This is then compared with the rotation angle achieved from the combination of two QWPs determined from eqn. \ref{achievable} about various axes $\bm{k}_{\bm{v}(\theta,\,\phi)}(\nu)$ for further analysis. The entire procedure is summarized in the  algorithm given below:\\
\\
\noindent \textbf{Algorithm}\\
\\
$\theta\in [0,\,\pi]$ \\
\phantom{$\theta$}\hspace{0.31cm} $\phi\in [0,\,2\pi]$
\vskip 0.2 cm
\noindent $\theta$ and $\phi$ denotes polar and azimuthal coordinates respectively, on the surface of Poincare sphere and with this input mid-plane $\bm{v}(\theta,\phi)$ is chosen. \vskip 0.002 cm
\phantom{$\theta$}\hspace{1.cm}$\nu_{0}\in [0,\,2\pi]$ \\
\phantom{$\theta$}\hspace{1.5cm}$\chi\in[0,\, \frac{\pi}{2}]$ 
\vskip 0.2 cm
\noindent Parameters $\nu_{0}$ and $\chi$ are used in eqns. \ref{Si}, \ref{Sf}  to obtain $\bm{S_{i}}$ and $\bm{S_{f}}$. By spanning $\nu_{0}$ and $\chi$, all possible $\bm{S_{i}}$ and $\bm{S_{f}}$ about the chosen mid-plane $\bm{v}(\theta,\phi)$ are obtained. Procedure is iterated about various mid-planes and in this way the transformation from every possible CPSL $\bm{S_{i}}$ to every other possible CPSL $\bm{S_{f}}$, is accounted.  \\
\\
Look for $\nu\in[0,\,2\pi]$, satisfying the condition:
\begin{center}
$\boxed{{\big(\cos\mu\,-\cos\mu_{q}\big)=0}}$ 
\end{center}
\vskip 0.002 cm
\noindent $\cos\mu_{q}$ is determined from eqn. \ref{achievable} and $\cos\mu$ is determined from eqn. \ref{Gen-mu-reqd}. Both are functions of the rotation axes parameter $\nu$. If parameter $\nu$ exists, satisfying the above boxed equation, then this ensures that there exists a rotation axis $\bm{k}_{\bm{v}(\theta,\phi)}(\nu)$ about which, the rotation angle $\mu$ required is same as the rotation angle $\mu_{q}$ achievable from two QWPs gadget. Hence, the transformation from $\bm{S_{i}}$ to $\bm{S_{f}}$ is possible using a combination of two QWPs. \\
\\
Further to find the orientation of two QWPs essential in transforming $\bm{S_{i}}$ to $\bm{S_{f}}$, the resulted solution $\nu$ is used to identify the rotation axis $\bm{k}_{\bm{v}(\theta,\phi)}(\nu)$ and the rotation angle $\mu_{q}$ achievable about it. From eqns. \ref{2qwp-CK} and \ref{qq-alpha-beta}, $\alpha,\, \beta$ are determined which is then used to identify the necessary orientations of two QWPs $\eta_{1},\, \eta_{2}$ in transforming $\bm{S_{i}}$ to $\bm{S_{f}}$.

\section{Conclusions}
\normalsize
\noindent We have formulated a fresh geometric approach for transforming every CPSL to every other CPSL and summarized it as in the algorithm. We have numerically verified that there always exists a rotation axis, about which the rotation angle required in transforming any two CPSL, can be achieved using two QWPs gadget. From this analysis, we conclude that the transformation from every CPSL to every other CPSL using only two QWPs is possible. Transformation of states with other degree of polarization is straight forward and it follows the procedure detailed for CPSL.

\section{Acknowledgments}
The author would like to thank G. Raghavan, S. Kanmani, K. Gururaj and
S. Sivakumar for their valuable suggestions.

\end{document}